\documentclass[aps,pra, twocolumn, floatfix, superscriptaddress]{revtex4-1}
\usepackage{amssymb, amsmath, bm, graphicx, color, xfrac} 
\DeclareGraphicsExtensions{.pdf,.png,.jpg}
\newcommand{\gvec}[1]{\boldsymbol{\mathrm{#1}}}
\begin{document}
\title{A route to non-Abelian quantum turbulence in spinor Bose--Einstein condensates}
\author{Thomas Mawson}
\affiliation{School of Physics, Monash University, Victoria 3800, Australia}
\author{Gary Ruben}
\affiliation{CSIRO Manufacturing Flagship, 343 Royal Parade, Parkville, Victoria 3052, Australia}
\author{Tapio Simula}
\affiliation{School of Physics, Monash University, Victoria 3800, Australia}
\date{\today}

\begin{abstract}
We have studied computationally the collision dynamics of spin-2 Bose--Einstein condensates initially confined in a triple-well trap. Depending on the intra- and inter-component relative phases of the initial state spinor wave function, the collision of the three condensate segments produces one of many possible vortex-antivortex lattices after which the system transitions to quantum turbulence. We find that the emerging vortex lattice structures can be described in terms of multi-wave interference. We show that the three-segment collisions can be used to systematically produce staggered vortex-antivortex honeycomb lattices of fractional-charge vortices, whose collision dynamics are known to be non-Abelian. Such condensate collider experiments could potentially be used as a controllable pathway to generating non-Abelian superfluid turbulence with networks of vortex rungs.
\end{abstract}
\maketitle


\section{Introduction}

For the past two decades, Bose--Einstein condensates (BECs) of dilute gases \cite{CornellWieman2002a,Ketterle2002a} have been a treasure trove of theoretical and experimental quantum physics. Amongst the gems discovered so far are superfluidity and integer-quantized vortices, the latter for single component BECs being characteristic of a scalar (spin-0) order parameter field \cite{Fetter2009a,Anderson2010a}. Such string-like quantized vortices puncture the order parameter field, defining its topology. The vorticity of the BEC superfluid is inherently connected to the configuration of the quantized vortices in the system. Vortices are motile and their dynamics can be regular, preserving the topology of the fluid, or chaotic, with topology-changing events mediated by vortex reconnections and vortex-antivortex pair-creation and annihilation processes. The latter type of behavior is prevalent in quantum or superfluid turbulence \cite{Feynman1955a,Barenghi2014a,TsubotaBook,Wilson2013a}.

Even more complex behaviors occur in spinor BECs, which are distinguished from the scalar BECs by their internal spin degree of freedom being unlocked \cite{Ho1998a,Ohmi1998a,StamperKurn2013a,Kawaguchi2012253}. The vector valuedness of the order parameter field of spinor superfluids greatly enriches the landscape of possible topological and magnetic structures in such systems \cite{Isoshima2002a,Makela2003a,Pogosov2005a,Kawaguchi2012a,Simula2012b}. In three-component (spin-1) spinor BECs, skyrmions \cite{Ruostekoski2001a,Khawaja2001a,Leslie2009a,Choi2012a,Lovegrove2014a}, monopoles \cite{Savage2003a,Pietila2009a,Ray2014a} and vortex sheets \cite{Kasamatsu2009a,Simula2011a} are among the topological structures \cite{VolovikBook} studied. The spin degree of freedom also allows these systems to possess turbulent spin currents \cite{Fujimoto2013a,Fujimoto2012b,Tsubota2013a} in addition to the turbulent mass currents. 

Five-component (spin-2) spinor BECs are predicted to exist in a variety of ground state phases \cite{Ciobanu2000a} and to host fractional-charge vortices \cite{Makela2003a,PhysRevLett.103.115301,kobayashi2011collision,Kobayashi2011b,Huhtamaki2009a}. Their so-called cyclic state in the polar phase is particularly interesting. Kobayashi \emph{et al.} showed that vortices in the cyclic phase ground state can be described by a non-Abelian algebra \cite{PhysRevLett.103.115301,kobayashi2011collision,Kobayashi2011b}. Collisions of such non-Abelian fractional vortices are topologically constrained and result in a rung joining the two resulting vortex lines, rather than the reconnection observed by conventional (Abelian) vortices in which the association of vortex line endings changes during the interaction. The non-Abelian character of the vortex collisions is therefore anticipated to result in a rung-turbulence if the system is driven far out of equilibrium \cite{PhysRevLett.103.115301,kobayashi2011collision,
Kobayashi2011b}. The non-Abelian fractional vortices in spin-2 BECs are also intrinsically interesting from the perspective of other constructs  such as topological quantum information processing \cite{RevModPhys.80.1083}, quantum field theories involving a spin-2 graviton \cite{Weinberg1965a} and non-Abelian cosmic string networks \cite{Copeland2007a,Copeland2010a}. However, in order to study  rung-mediated non-Abelian quantum turbulence experimentally, a method to controllably produce the novel non-commutative types of vortices is required.

\begin{figure}
\centering
\includegraphics[width=\columnwidth]{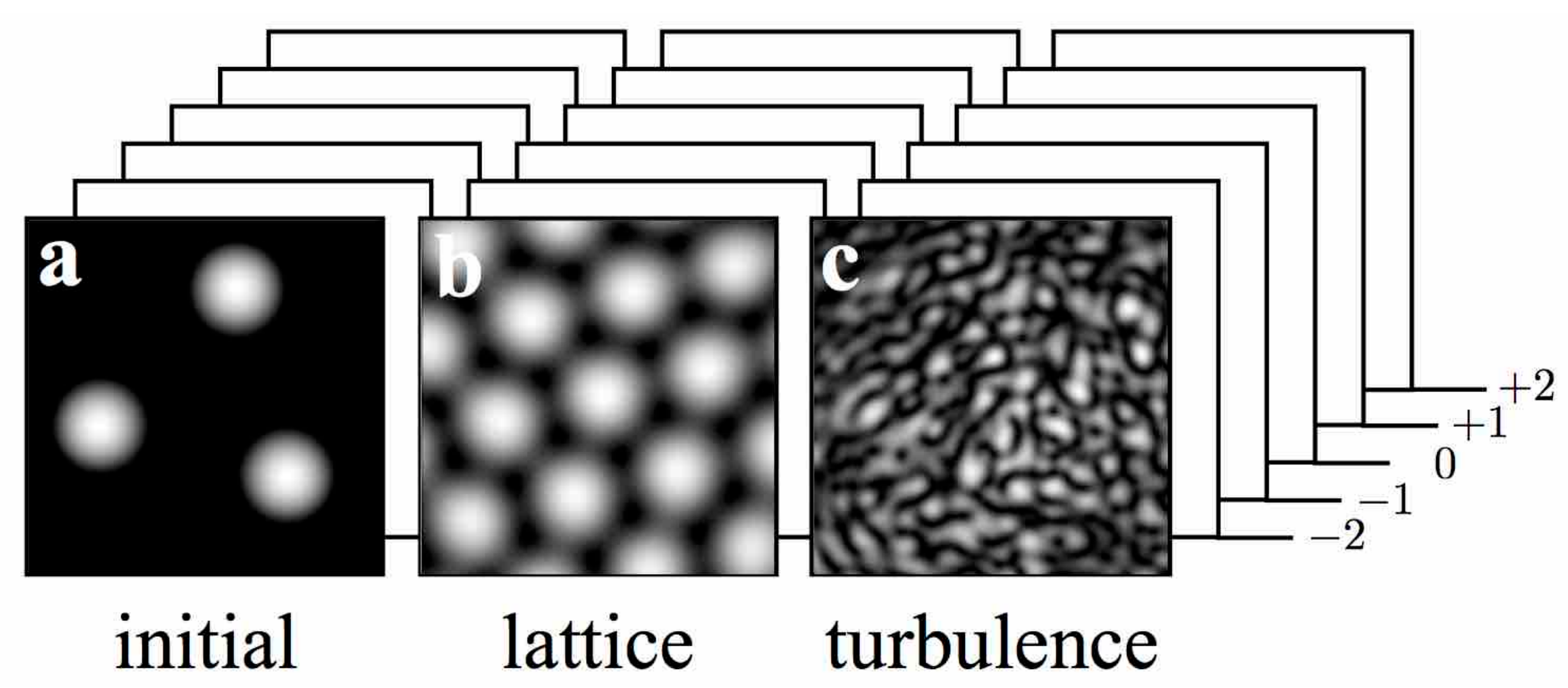}
\caption{Schematic of our numerical experiment showing the total particle density at three different times of evolution. (a) The five hyperfine states behind the total density images of the $F=2$ Bose--Einstein condensate are initially confined in a species-independent triple-well trap, superposed with a global harmonic potential. (b) Upon removal of the triple-well potential, the three condensate fragments expand and interfere in the central portion of the harmonic trap. The multi-wave interference of the condensate fragments produces transient honeycomb vortex-antivortex lattices in each of the occupied hyperfine spin states. (c) At later times the system transitions to quantum turbulence.}
\end{figure}

Several techniques exist for producing quantized vortices in BECs \cite{Anderson2010a} including nucleation by rotating traps \cite{Penckwitt2002a,Simula2002a,Kasamatsu2003a}. Here we focus on vortex production based on multi-wave interference to achieve a controllable and repeatable technique to generate quantum turbulence. Indeed, interference of three or more waves can produce lattices of quantized vortices and antivortices \cite{Nicholls1987a,Ruben2007a,Ruben2008a,Ruben2010a,Simula2011a}. Scherer et al. used such a method and by colliding three Bose--Einstein condensate fragments they observed quantized vortices in the system \cite{Scherer2007a}. In their experiment an external potential was used for separating the condensate initially into three condensate fragments. Adiabatic removal of the separating potential provided a statistical prediction of the presence or absence of a vortex in the resulting condensate, depending on the random relative phases of the initial condensate components. In contrast with this experiment, under sufficiently rapid non-adiabatic removal of the separating potential, three colliding condensate 
fragments have been predicted and demonstrated to form a honeycomb vortex-antivortex lattice, equivalent to two interleaved Abrikosov lattices---one of vortices and the other of anti-vortices \cite{Gonzales2008a,Ruben2008a,henderson2009experimental}. Similar honeycomb vortex lattices could also be produced by using aberrated matter wave lensing technique \cite{Petersen2013a,Simula2013a}. Moreover, Kj\ae rgaard group has developed a versatile optical tweezer collider for cold atoms \cite{Roberts2014a}, which could be extended to two-dimensional collision geometry to achieve generic multi-wave condensate collisions with controllable initial momentum vectors of the wave packets. For a two-component pseudo-spin system, three-wave collisions lead to condensate pseudo-spin textures \cite{Ruben2010a}.

In the remainder of this manuscript, we extend the three-wave interference concept to an $F=2$ spinor BEC. In particular, we show that it is possible to use such a multi-wave interference technique to create vortex states that host non-Abelian fractional vortices which are anticipated to lead to novel rung-mediated quantum turbulence. In Sec. II we outline an analytical method for determining the axisymmetric vortex types that form in the aforementioned condensate collisions. In Sec. III we describe the vortex lattices composed of these vortex types. The simulation results presented in Sec. IV include regimes of linear dynamics which results in the nucleation of vortex lattices and the non-linear dynamics which is anticipated to lead to non-Abelian rung-mediated quantum turbulence. We discuss the obtained results in terms of three key observables; the total condensate particle density, magnetization density and the spin singlet pair amplitude density.


\section{Axisymmetric quantized vortices in an $F = 2$ condensate}

In the condensate collision experiment under consideration the relevant dynamics are examined within a transverse plane of the BEC, which treats vortex phase singularities as 0-dimensional point objects in the 2d plane, equivalent to infinitely extended 1d parallel nodal lines in the 3d complex scalar field. This simplification means that it is reasonable to restrict the discussion to vortices whose cores are aligned with the quantization axis perpendicular to the plane of the triple-well potential. Motivated by this observation, we here restrict the discussion to 2d systems unless otherwise stated.

\begin{figure}
\centering
\includegraphics[width=0.65\columnwidth]{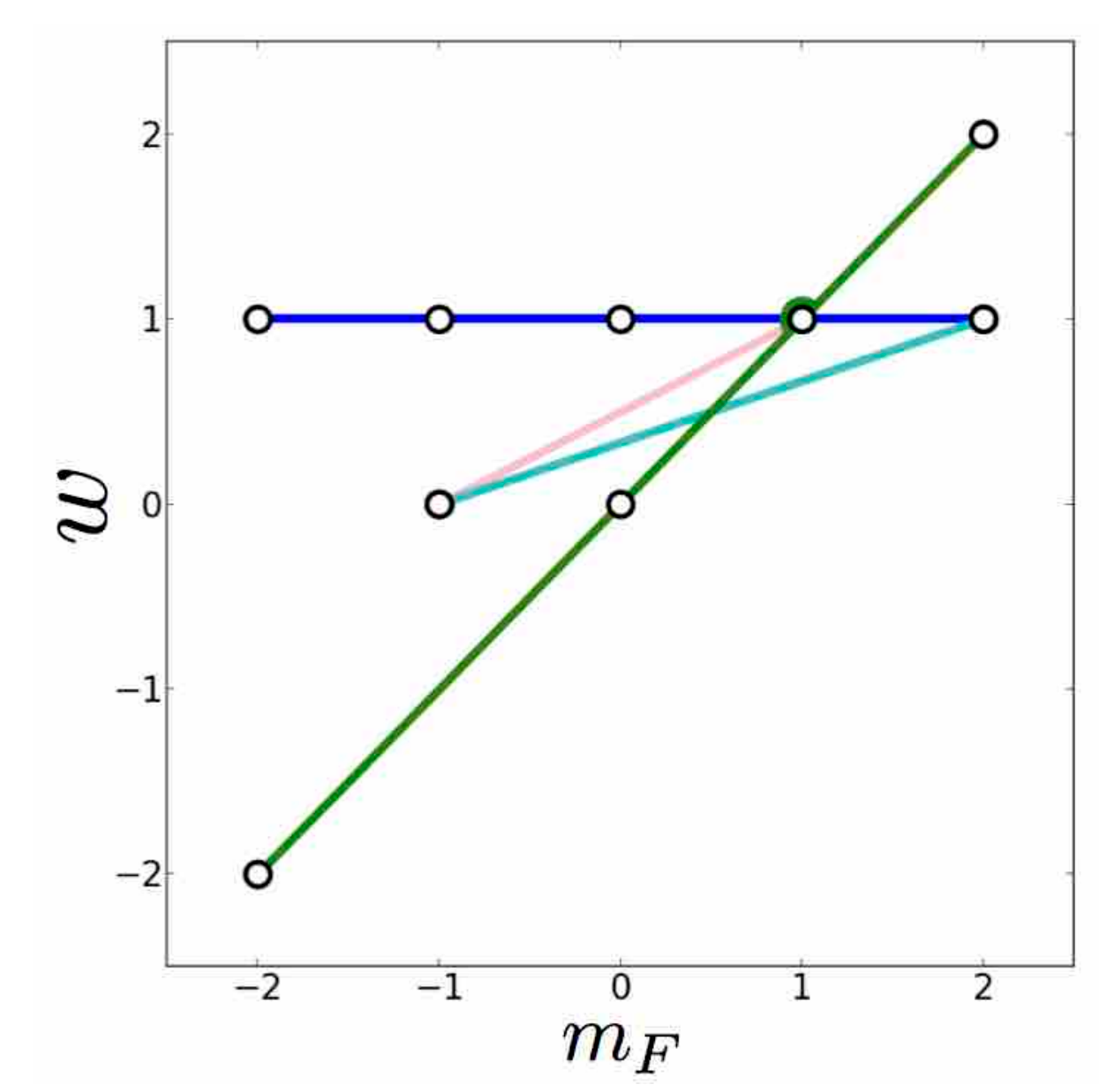}
\caption{(Color online) Selection of axisymmetric vortices in an $F=2$ spinor condensate represented in terms of the winding numbers $w$ of each hyperfine spin state $m_F$ of the condensate. Circles mark hyperfine spin components with nonzero condensate population. In this graphical notation each vortex corresponds to a straight line.}
\label{fig2}
\end{figure}

\subsection{Axisymmetric vortices}
Without a loss of generality, we may express the five-component order parameter of the $F = 2$ spinor Bose--Einstein condensate
in a Madelung form
\begin{equation}
\Psi\left(\gvec{r},\,t\right) = \begin{pmatrix} \left|\psi_2\left(\gvec{r},\,t\right)\right|e^{iS_2\left(\gvec{r},\,t\right)} \\ \left|\psi_1\left(\gvec{r},\,t\right)\right|e^{iS_1\left(\gvec{r},\,t\right)} \\ \left|\psi_0\left(\gvec{r},\,t\right)\right|e^{iS_0\left(\gvec{r},\,t\right)} \\ \left|\psi_{-1}\left(\gvec{r},\,t\right)\right|e^{iS_{-1}\left(\gvec{r},\,t\right)} \\ \left|\psi_{-2}\left(\gvec{r},\,t\right)\right|e^{iS_{-2}\left(\gvec{r},\,t\right)}  \end{pmatrix},
\label{eq1}
\end{equation}
where $\left|\psi_j\left(\gvec{r},\,t\right)\right|$ and $S_j\left(\gvec{r},\,t\right)$ are, respectively, the amplitude and the phase functions of the $j$th hyperfine spin component of the spinor wavefunction $\Psi\left(\gvec{r},\,t\right)$. A constraint of axisymmetric superflow mass current  around the vortex phase singularities may be imposed by specifying the spatial phase functions to be of the form
\begin{equation}
S_j\left(\gvec{r}\right) = w_j\theta,
\end{equation}
where $w_j$ is the integer winding number of a phase singularity in the $j$th spin state and $\theta$ is the polar phase angle. Positive and negative values of $w_j$ in single component condensates correspond to vortices and antivortices, respectively. Furthermore, it is convenient to use a winding number notation whereby a 5-tuple $\left(w_{-2},\, w_{-1},\, w_0,\, w_{1},\, w_{2}\right)$ can be used to refer to the axisymmetric vortices.  Here, the subscript refers to the hyperfine state $m_F$. A spin state unpopulated by atoms is denoted with a $\times$ marker. Different winding number tuples describe different vortex species each with particular orbital and spin angular momentum quantum numbers $\ell$ and $s$, respectively.

The phase functions $S_j$ of a spinor $\Psi\left(\gvec{r}\right)$ can be unwound by a transformation
\begin{equation}
\Psi_0\left(\gvec{r}\right)=e^{-i\gamma}e^{i\eta_x f_x}e^{i\eta_y f_y}e^{i\eta_z f_z} \Psi\left(\gvec{r}\right),
\label{gaugespin}
\end{equation}
where $f_i$ are Pauli spin-matrices for a particle with spin $f$, and $\gamma$ and $\eta_i$ are respectively the gauge and spin rotation angles of the transformation. The axisymmetric vortices only involve spin rotation about the chosen $z$ quantization axis, such that $\eta_x=\eta_y=0$. This constrains the allowed winding numbers $w_j$ of the nearest populated hyperfine spin states $j$ and $j+1$ such that $w_{j+1}-w_j = n$, where $n$ is an integer. This constraint allows a graphical representation of the axisymmetric vortices shown in Fig.~\ref{fig2} as a straight line in the plane spanned by the hyperfine spin quantum number $m_F=j$ and the winding number $w$ \cite{Isoshima2002a,Pogosov2005a}. We will further refer to such axisymmetric vortices by a pair of numbers $\left(\ell,\,s\right)$ where $\ell=\gamma$ and $s=\eta_z$ quantum numbers are related to the orbital angular momentum and spin angular momentum of the vortex, respectively.

\begin{table}
\caption{Selection of axisymmetric vortices in the $F=2$ spinor condensate. We refer to the vortices either using the spin and orbital angular momentum quantum numbers $\left(\ell,s\right)$, or the winding number tuple. A winding number of $\times$ indicates a zero population of particles in that spin state. Vortices indexed with $\ast$ belong to one of the four non-Abelian vortex classes.}
\label{table_vortex_selection}
\begin{tabular}[t]{lc}
\hline\hline
 $\left(\ell,\,s\right)$ & winding number tuple \\ \hline
 $\hspace{6pt}\left(1,\,0\right)$ & $\left(\times,\,\times,\,1,\,\times,\,\times\right)$ \\
 $\hspace{6pt}\left(-1,\,0\right)$ & $\left(\times,\,\times,\,-1,\,\times,\,\times\right)$ \\
 $\ast\left(-\frac{2}{3},\,\frac{1}{3}\right)$ & $\left(\times,\,-1,\,\times,\,\times,\,0\right)$ \\
 $\hspace{6pt}\left(\frac{1}{2},\,\frac{1}{2}\right)$ & $\left(\times,\,0,\,\times,\,1,\,\times\right)$ \\
 $\hspace{6pt}\left(-\frac{1}{2},\,\frac{1}{2}\right)$ & $\left(\times,\,-1,\,\times,\,0,\,\times\right)$ \\
 $\hspace{6pt}\left(\frac{1}{2},\,-\frac{1}{4}\right)$ & $\left(1,\,\times,\,\times,\,\times,\,0\right)$ \\
 $\hspace{6pt}\left(-\frac{1}{2},\,-\frac{1}{4}\right)$ & $\left(0,\,\times,\,\times,\,\times,\,-1\right)$\\
 $\hspace{6pt}\left(\frac{1}{3},\,-\frac{2}{3}\right)$ & $\left(\times,\,1,\,\times,\,\times,\,-1\right)$ \\
 $\ast\left(\frac{1}{3},\,\frac{1}{3}\right)$ & $\left(\times,\,0,\,\times,\,\times\,,1\right)$ \\
 $\ast\left(0,\,1\right)$ & $\left(-2,\,\times,\,0,\,\times,\,2\right)$ \\
 $\hspace{6pt}\left(0,\,-1\right)$ & $\left(\times,\,1,\,\times,\,-1,\,\times\right)$ \\
 $\hspace{6pt}\left(0,\,\frac{1}{2}\right)$ & $\left(-1,\,\times,\,0,\,\times,\,1\right)$ \\
 
 \hline \hline
\end{tabular}
\end{table}

\subsection{Vortex structures}
\label{section_core_structure}

In Table \ref{table_vortex_selection} we have listed a few axisymmetric vortex types. The vortex $\left(1,\,0\right)$, is a scalar vortex with pure mass circulation ($\ell =1$) and vanishing spin current ($s=0$). A complementary example is the vortex $\left(0,\,1\right)$ which has pure spin current and no mass current. In addition to these integer vortices, a generic fractional vortex may have a mixture of mass and spin currents, in which case the values of $\ell$ and $s$ can be rational numbers. The $\left(\frac{1}{2},\,\frac{1}{2}\right)$ vortex has a half quantum of both mass and spin winding and is the spin-2 analog of the half-quantum vortex \cite{JETP72}. Three further examples of fractional vortices are $\left(\frac{1}{2},\,-\frac{1}{4}\right)$, $\left(\frac{1}{3},\,\frac{1}{3}\right)$ and $\left(-\frac{2}{3},\,\frac{1}{3}\right)$, with the latter two existing in the cyclic ground state of the $F=2$ condensate.

The spin degree of freedom allows for a variety of vortex core structures to exist \cite{Isoshima2002a,Pogosov2005a,Kobayashi2012a}. The vortex states present in scalar (spin-0) condensates have vanishing condensate particle density in the vortex core, although they can be partially filled by quantum depletion and thermal atoms \cite{Isoshima1999a,Virtanen2001a,Virtanen2001b,Virtanen2001c}. In contrast, in spinor condensates the void left by a vortex in one spin component may be filled by condensate particles in other spin components. The core structure of a generic vortex in $F=2$ condensate may be conveniently characterized in terms of three functions. These are
the total particle density \cite{Kawaguchi2012a}
\begin{equation}
n(\gvec{r}) =  |\psi_{2}|^2 + |\psi_{1}|^2 + |\psi_{0}|^2  + |\psi_{-1}|^2 +|\psi_{-2}|^2
\label{ndens}
\end{equation}
the  magnetization density
\begin{equation}
 M(\gvec{r}) = 2|\psi_2|^2+ |\psi_1|^2 - |\psi_{-1}|^2 -2|\psi_{-2}|^2
\label{mdens}
\end{equation}
and the spin singlet pair amplitude 
\begin{equation}
 A(\gvec{r}) = \frac{1}{\sqrt{5}}\left(2\psi_{2}\psi_{-2}\right)-2\psi_{1}\psi_{-1}+\psi^2_0.
\label{adens}
\end{equation}

\subsection{Non-Abelian vortices}

Each spinor vortex can be associated with a topological charge which satisfies either Abelian or non-Abelian algebra \cite{kobayashi2011collision}. If the multiplication of two topological charges is non-commutative then the vortices and their interactions are described as non-Abelian. Such non-Abelian behavior is manifest in their collision dynamics. When two non-Abelian vortices collide it is topologically forbidden for them to undergo a reconnection. Instead, a rung vortex emerges at the interaction site bridging the two vortex lines \cite{kobayashi2011collision}. The cyclic ground state of the $F = 2$ condensate is symmetric under rotations in the non-Abelian tetrahedral group. Vortices formed in the cyclic ground state of such spinor condensates have a topological charge corresponding to one of these group elements and such vortices inherit the non-Abelian property of the tetrahedral group. It is therefore interesting to investigate if such non-Abelian vortices could be generated experimentally in 
spin-2 BECs for studies of rung-mediated quantum turbulence. The non-Abelian vortices relevant to this work are marked in Table \ref{table_vortex_selection} by an asterisk.

\section{Three-wave interference yields honeycomb vortex-antivortex lattices}
The destructive interference of two wave packets produces dark stripe solitons, each of which may subsequently disintegrate due to nonlinear interactions into rows of alternating vortices and antivortices \cite{Burger1999a,Denschlag2000a,Dutton2001a,Cetoli2013a}. However, the destructive interference of three waves may produce lattices of vortices and antivortices in predictable regular honeycomb lattice structures \cite{Nicholls1987a,Ruben2007a,Ruben2008a,Ruben2010a,Simula2011a}. Armed with this insight, we will first consider a collision of three single-component scalar condensates and thereafter extend the results to the multi-component spinor condensates. Our aim is to use three-wave interference technique to deterministically produce vortices and antivortices of non-Abelian kind to ignite rung-mediated quantum turbulence.

\subsection{Collision of three scalar condensates}

In the absence of external potentials and magnetic fields the Hamiltonian density of the $F=2$ condensate may be expressed as \cite{Kawaguchi2012a}
\begin{equation}
 H = T(\gvec{r}) + \frac{c_0}{2}n^2(\gvec{r}) + \frac{c_1}{2}\left|\gvec{F}(\gvec{r})\right|^2 + \frac{c_2}{2}\left|A(\bf r)\right|^2,
\end{equation}
where the single particle term $T(\gvec{r})=-\frac{\hbar^2}{2m}\sum_j \psi^*_j (\gvec{r}) \nabla^2\psi_j (\gvec{r})$, the spin vector 
$\gvec{F}(\gvec{r})=\sum_{ij} \psi^*_i(\gvec{r})\gvec{f}_{ij} \psi_{j}(\gvec{r})$, where
$\gvec{f}_{ij}$ are the elements of the spin-2 Pauli matrices, and the total particle density $n(\gvec{r})$ and spin singlet pair amplitude $A(\gvec{r})$ were defined in Sec.~\ref{section_core_structure} and $c_i$ are coupling constants.

If the kinetic energy of the three colliding condensate components is much larger than any of the interaction terms the system may be considered to be weakly non-linear and we may set $c_i=0$. Under such circumstances the resulting spinor wavefunction on short time-scales can be estimated via the superposition principle of linear waves. While the condensate fragments are modeled as three symmetrically arranged wave packets of equal initial population and shape, on a local scale their interference may be modelled as that of three plane waves. This approximation is represented by a wavefunction  
\begin{equation}
 \psi\left(\gvec{r}\right) = \sum^3_{j=1}e^{i\left({\gvec{k}_j\cdot\gvec{r} + \phi_j}\right)},
\end{equation}
where $\gvec{k}_j$ are the momentum vectors of equal magnitude $|\gvec{k}|$ of the three colliding condensate fragments and $\phi_j$ specifies the phase of the $j$th condensate fragment relative to $\phi_1=0$. 

\begin{figure}[t]
\centering
 \includegraphics[width=\columnwidth]{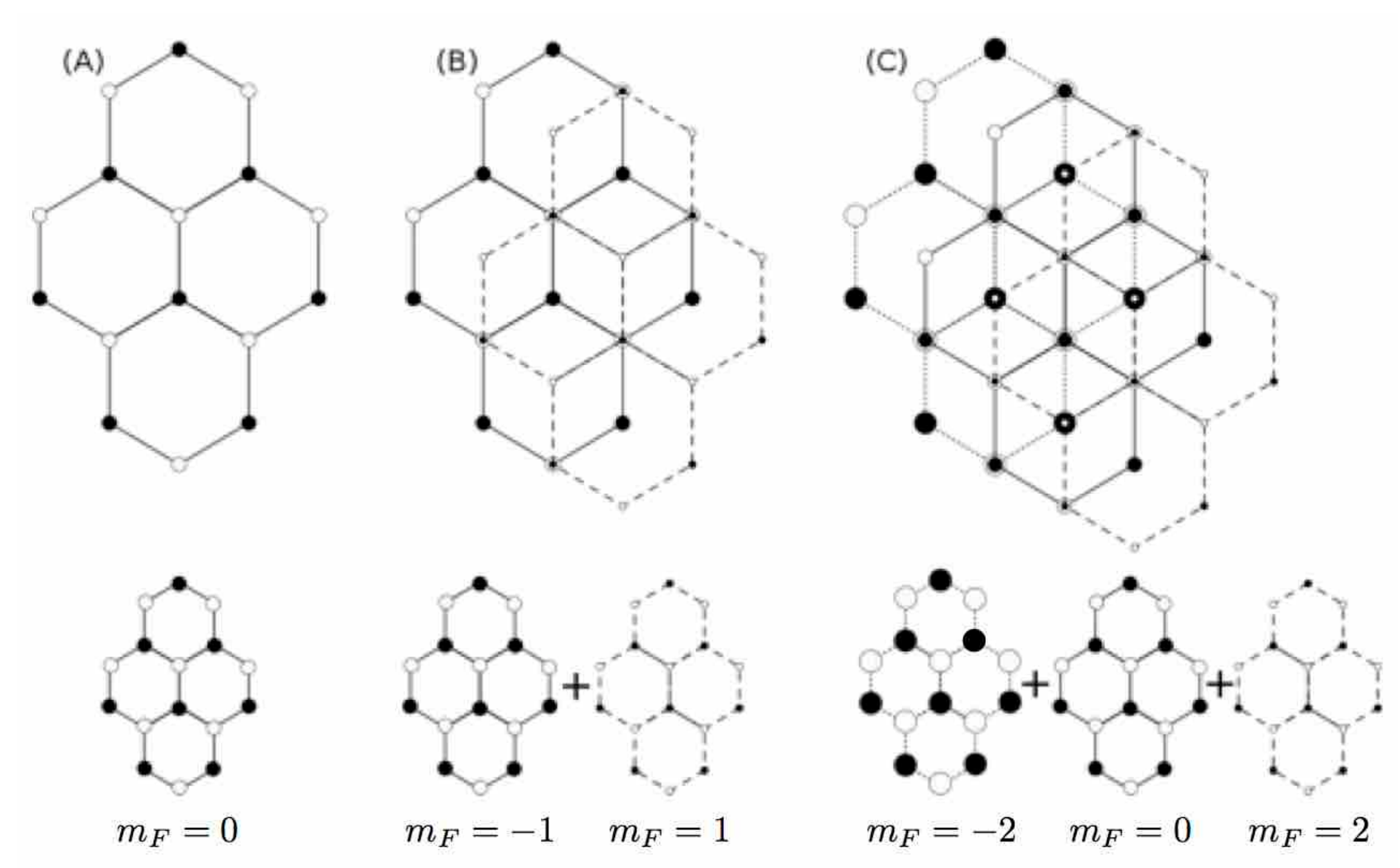}
 \caption{Structure of vortex lattices with black and white circles corresponding to the locations of vortices and antivortices, respectively. (a) A honeycomb lattice structure due to three-plane-wave interference in a single spin state of a condensate. (b) AB stacking of two honeycomb lattices. (c) ABC stacking of three honeycomb lattices.}
\label{graphene}
\end{figure}

The quantized vortices are nodal lines of the complex valued wavefunction and are identified as singularities of the phase function $\arg(\psi)$. For destructive three-plane-wave interference the locations of the vortices, $r^+_{n,m}$, and antivortices, $r^-_{n,m}$, are given by the simple geometric relations
\begin{align}
r^+_{n,m} =&  \frac{4\pi}{3|\gvec{k}|}\left(n+m/2+1/2\pi\left(\phi_2+\phi_3\right)\right)\ \hat{\gvec{e}}_x \notag \\ &+ \frac{2\pi}{\sqrt{3}|\gvec{k}|}\left(m+\phi_2/\pi\right) \hat{\gvec{e}}_y,\notag\\
r^-_{n,m} =&  \frac{4\pi}{3|\gvec{k}|}\left(n+m/2+1/2\pi\left(\phi_2+\phi_3\right)+ 1/2\right)\ \hat{\gvec{e}}_x \notag \\ &+ \frac{2\pi}{\sqrt{3}|\gvec{k}|}\left(m +\phi_2/\pi +1/3\right) \hat{\gvec{e}}_y
\label{planemodel}
\end{align}
describing a honeycomb lattice, as illustrated in Fig.~\ref{graphene}(a). The $\hat{\gvec{e}}_i$ are Cartesian basis vectors.
In comparison, by treating the colliding wave packets as Gaussian functions the interference is described by the wavefunction \cite{bransden2000introduction,Ruben2008a}
\begin{align}
 \psi\left(\gvec{r},\,t\right) =& \sqrt{\frac{\pi^{-\frac{1}{2}}\frac{\Delta p}{\hbar}}{1+i\frac{\left(\Delta p\right)^2t}{m\hbar}}} \notag \\ & \times \sum^3_{j=1}\exp\left(\frac{-\left(\frac{\Delta p}{\hbar}\right)^2\left|\gvec{r}-\gvec{r}_j\right|^2}{2\left[1+i\frac{\left(\Delta p\right)^2t}{m\hbar}\right]}+i\phi_j\right),
\end{align}
where $\Delta p$ defines the momentum uncertainty or the width of the initial condensate fragments and $\left| \gvec{r}-\gvec{r}_j \right | $ is the distance between the center of the $j$th Gaussian and an observation point. 
The subsequent vortex and antivortex locations, derived in detail by Ruben et al. \cite{Ruben2008a}, are respectively 
\begin{align}
{r^\prime}^+_{n,m} =&  \frac{1}{2}\left(r-\frac{\beta M\left(m\right)}{r}\right)\ \hat{\gvec{e}}_x \notag \\
& +\frac{1}{2\sqrt{3}}\left(r-\frac{\beta\left[2N\left(n\right)- M\left(m\right)\right]}{r}\right)  \hat{\gvec{e}}_y, \notag\\
{r^\prime}^-_{n,m} =&  \frac{1}{2}\left(r-\frac{\beta M'\left(n\right)}{r}\right)\ \hat{\gvec{e}}_x \notag \\
& +\frac{1}{2\sqrt{3}}\left(r-\frac{\beta\left[2N'\left(m\right)- M'\left(n\right)\right]}{r}\right)  \hat{\gvec{e}}_y,
\label{Rubenmodel}
\end{align}
where $r$ is the separation between the centers of each condensate fragment and $\beta = 2\left[\left(\hbar t\right)^2+m^2\left(\hbar/\Delta p\right)^4\right]/3m\hbar t$. The integers $n$ and $m$ index the lattice points via the functions 
\begin{align}
 M\left(m\right) &= 2\pi\left[1+3\left(m-\frac{\phi_2}{2\pi}\right)\right] \notag \\
 N\left(n\right) &= 2\pi\left[2+3\left(n-\frac{\phi_3}{2\pi}\right)\right] \notag \\ 
 M'\left(n\right) &= 2\pi\left[2+3\left(n-\frac{\phi_2}{2\pi}\right)\right] \notag \\
 N'\left(m\right) &= 2\pi\left[1+3\left(m-\frac{\phi_3}{2\pi}\right)\right]. \notag \\ 
\end{align} 

Interestingly, also the Gaussian wave packet model, Eq.~\ref{Rubenmodel}, produces a uniform honeycomb vortex lattice and hence the vortex lattice vectors of Eq.~(\ref{planemodel}) and Eq.~(\ref{Rubenmodel}) can be mapped onto each other at any time. Here we use destructive three-plane-wave interference as an analytical model for comparison with the numerical results.\\
 
\subsection{Collision of three spinor condensates}

Using the three-plane-wave linear superposition, as outlined for the scalar condensate, the spinor wave function describing the lattice structure local to the trap center is 
\begin{equation}
 \Psi\left(\gvec{r}\right) = \sum^3_{j=1}\begin{pmatrix} e^{i\left(\gvec{k}_j\cdot\gvec{r} + \phi_{2j}\right)} \\ e^{i\left(\gvec{k}_j\cdot\gvec{r} + \phi_{1j}\right)} \\ e^{i\left(\gvec{k}_j\cdot\gvec{r} + \phi_{0j}\right)} \\ e^{i\left(\gvec{k}_j\cdot\gvec{r} + \phi_{-1j}\right)} \\ e^{i\left(\gvec{k}_j\cdot\gvec{r} + \phi_{-2j}\right)}  \end{pmatrix},
\label{analyticalspinor}
\end{equation}
where $\left|\psi_i\left(\gvec{r}\right)\right|$ is constant and $\phi_{ij}$ specifies the relative phase of the $j$th condensate fragment in the $i$th hyperfine spin component. The magnitude of the momentum vector $\gvec{k}_j$ determines the vortex lattice spacing and is used as a free parameter for matching the semi-analytical description and the numerical results.

Within the weak interaction approximation of three-source interference of spinor condensates, the lattice structure formed in each hyperfine spin state is independent of influence from the condensate particles in other spin states. The lattice geometry in each spin state is thus exactly the same as in the analytical model for a scalar condensate. At every point in space, each condensate spin component $\psi_j (\gvec{r})$ may have a vortex, an antivortex or be vortex free. The vortex lattices in different spin components can be formed with any desired alignment \cite{Ruben2010a}, leading to various different stackings analogous to multilayer graphene structures \cite{Koshino2013a}. 

Figure \ref{graphene}(a) shows a honeycomb vortex lattice structure in a single component scalar (spin-0) condensate. Frame (b) shows an AB stacking of two honeycomb lattices and frame (c) shows an ABC stacking of three-layer honeycomb lattices. Both (b) and (c) yield three different kinds of axisymmetric vortices in the 
global order parameter and the same is true for ABA and BAB stackings. Unshifted
stackings such as AA and AAA all result in two different kinds of vortices only, which are equivalent to the scalar condensate vortices and antivortices shown in (a). Populating 4 or all 5 spin components further increases the complexity and the variety of possible vortex states.
A list of interesting initial states and their corresponding lattice vortices is presented in Table~\ref{table_initial_states}. In all cases, the resulting lattice has a net zero winding number. This follows since the honeycomb lattice in each populated component has equal numbers of vortices and antivortices and hence a net-zero winding number for each component.

\begin{table}
\caption{A list of initial vortex states, corresponding to those defined in Sec. II, along with the vortices present in the lattice after the three source interference.}
\label{table_initial_states}

\begin{tabular}[b]{lcc}
 \hline
 \hline
 Initial vortex & lattice vortices \\ \hline
 $\left(\frac{1}{2},\,\frac{1}{2}\right)$ &  $\left(-\frac{1}{2},\,\frac{1}{2}\right)$, $\,\left(\frac{1}{2},\,\frac{1}{2}\right)$, $\,\left(0,\,-1\right)$\\
 $\left(0,\,1\right)$                     &  $\left(\frac{1}{2},\,-\frac{1}{4}\right)$ +$\,\left(-1,\,0\right)$,$\,\left(-\frac{1}{2},\,-\frac{1}{4}\right)$ + $\left(1,\,0\right)$,$\,\left(0,\,\frac{1}{2}\right)$\\
 $\left(\frac{1}{3},\,\frac{1}{3}\right)$ &  $\left(\frac{1}{3},\,\frac{1}{3}\right)$,$\,\left(-\frac{2}{3},\,\frac{1}{3}\right)$ ,$\,\left(\frac{1}{3},\,-\frac{2}{3}\right)$\\\\
\hline\hline
\end{tabular}%
\end{table}

\subsection{Generating non-Abelian lattice vortices}

\begin{figure}[!ht]
 \centering
 \includegraphics[width=\columnwidth]{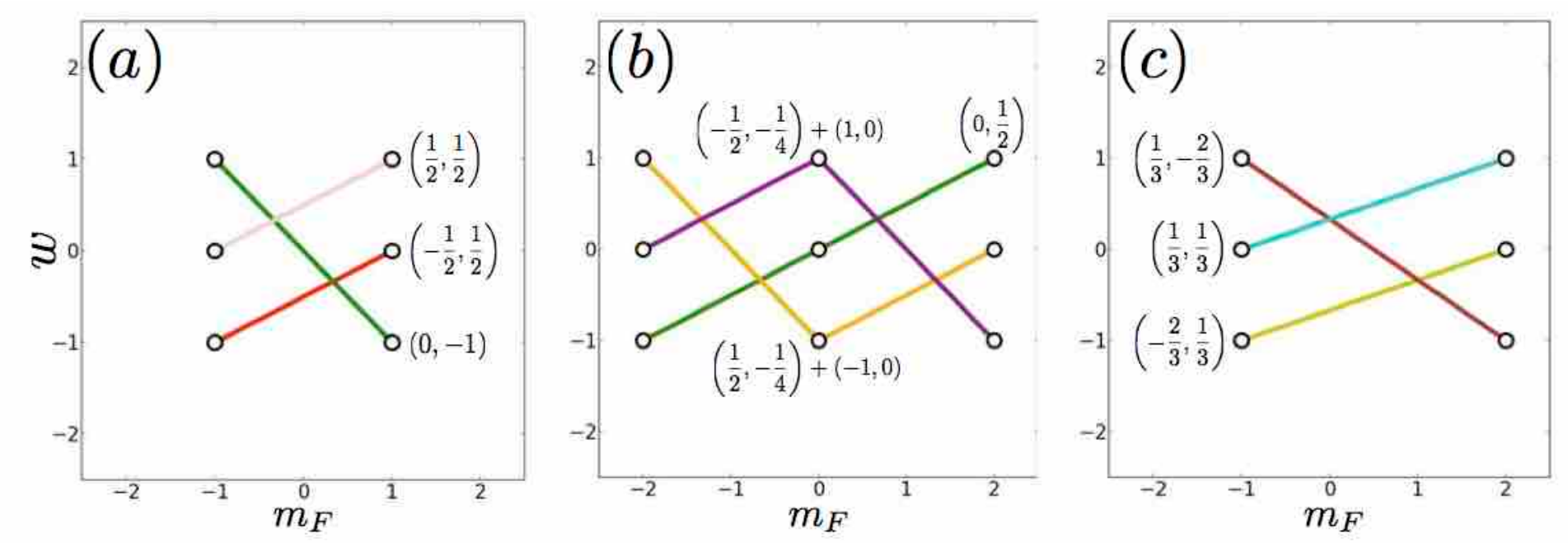}
 \caption{(Color online) The lattice vortices of a selection of initial states represented in terms of the winding numbers $w$ of each hyperfine spin state $m_F$ of the condensate. Circles mark hyperfine spin components with nonzero condensate population. In this graphical notation each axisymmetric vortex corresponds to a set of nodes for each populated spin state joined by a straight line, and is labeled using the naming convention listed in Table~\ref{table_vortex_selection}. (a)-(c) show the vortices that emerge in the lattice when using an initial state vortex $\left(\frac{1}{2},\,\frac{1}{2}\right)$, $\left(0,\,1\right)$ and $\left(\frac{1}{3},\,\frac{1}{3}\right)$, respectively.}
  \label{K9}
\end{figure}

For collisions of condensates with population in two hyperfine spin states the $\left(\frac{1}{3},\,\frac{1}{3}\right)$ vortex corresponds to one of the possible initial configurations and in the semi-analytical model is achieved by setting $\phi_{-1j}={\rm 0}$ and $\phi_{2j}=0,\,\frac{2\pi}{3},\,\frac{4\pi}{3}$ for $j=1,\,2,\,3$, respectively, with the other spin components left empty. The $\left(\frac{1}{3},\,\frac{1}{3}\right)$ is also one of the non-Abelian vortex types of the cyclic ground state. The lattice vortices produced in this case are $\left(\frac{1}{3},\,\frac{1}{3}\right)$, $\left(-\frac{2}{3},\,\frac{1}{3}\right)$  and $\left(\frac{1}{3},\,-\frac{2}{3}\right)$ as detailed in Tables \ref{table_vortex_selection} and \ref{table_initial_states} and Fig.~\ref{K9}(c). The $\left(\frac{1}{3},\,-\frac{2}{3}\right)$ vortex is unstable in the cyclic phase \cite{Kobayashi2011b}. However, collisions between $\left(\frac{1}{3},\,\frac{1}{3}\right)$ and $\left(-\frac{2}{3},\,\frac{1}{3}\right)$  vortices are expected to be 
non-Abelian resulting in rung formation \cite{Kobayashi2011b}. 
Therefore, as the vortex lattice produced by the $\left(\frac{1}{3},\,\frac{1}{3}\right)$ initial state breaks up and enters the nonlinear regime in its evolution, it is anticipated that new rung-mediated quantum turbulence may be initiated this way \cite{kobayashi2011collision}. 

\section{Gross-Pitaevskii dynamics}

To confirm the validity of the semi-analytical model for vortex lattice generation when particle interactions are accounted for, we have simulated the collisions of three $F = 2$ condensate fragments using the mean-field theory. The system is modeled in two dimensions using the five-component $F = 2$ spinor Gross-Pitaevskii equations \cite{Kawaguchi2012253}
\begin{align}
  i\hbar\partial_{t}\psi_{\pm 2} &= \left(-\frac{\hbar^2 \nabla^2}{2m}+V_\textrm{ext}+c_{0}n\pm2c_{1}F_{z}-\mu\right)\psi_{\pm2}\notag \\
&+c_{1}F_{\pm}\psi_{\pm1}+\frac{c_2}{\sqrt{5}}A\psi^*_{\mp2}, \notag\\
  i\hbar\partial_{t}\psi_{\pm 1} &= \left(-\frac{\hbar^2 \nabla^2}{2m}+V_\textrm{ext}+c_{0}n\pm 
  c_{1}F_{z}-\mu\right)\psi_{\pm1} \notag\\
  &+c_{1}\left(\frac{\sqrt{6}}{2}F_{\mp}\psi_{0}+F_{\pm}\psi_{\pm2}\right)-\frac{c_2}{\sqrt{5}}A\psi^*_{\mp1}, \notag\\
  i\hbar\partial_{t}\psi_{0} &= \left(-\frac{\hbar^2 \nabla^2}{2m}+V_\textrm{ext}+c_{0}n-\mu\right)\psi_{0} \notag\\
  &+c_{1}\frac{\sqrt{6}}{2}\left(F_{+}\psi_{1}+F_{-}\psi_{-1}\right)+\frac{c_2}{\sqrt{5}}A\psi^*_{0},
 \end{align}
which govern the dynamics of the spinor condensate. Here $\psi_{i}$ for $i=(0,\,\pm1,\,\pm2)$ denotes a single hyperfine spin component of the spinor and $F_{\pm}$ are the angular momentum raising and lowering operators. The terms $c_0,\,c_1$ and $c_2$ are the effective coupling constants and $\mu$ is the chemical potential. The coupling constants depend on the scattering lengths of the particle interaction channels which for ${}^{87}$Rb  are $a_0=87.4a_{\rm B}$, $a_2=92.4a_{\rm B}$ and $a_4=100.5a_{\rm B}$ \cite{PhysRevA.64.053602} in units of the Bohr radius $a_{\rm B}$. The number of particles in the system is $N=5.0 \times 10^4$. The simulations are performed on a Cartesian numerical grid with $2048\times2048$ points using the XMDS2 \cite{dennis2013xmds2} differential equation solving package.

The trapping potential is a combination of a harmonic oscillator potential with a frequency $\omega = 2\pi\times 4.9$ Hz and a sum of three localized Gaussian terms. This produces a triple-well potential, $V_\textrm{ext}=m\omega^2r^2/2+\sum_{i=1}^3 P\exp\left(-\frac{1}{2}({\bf r}-{\bf r}_i)^2/\sigma^2 \right)$, where  $P=300\;\hbar\omega$ and $\sigma=18.8\; a_\textrm{osc}$ are the potential height and standard deviation of the Gaussian potentials centered at positions ${\bf r}_i$. The unit of distance $a_\textrm{osc}=\sqrt{\hbar/2m\omega}$. The initial state densities are produced numerically using imaginary time propagation and the initial-state spinor is chosen to represent one of the possible axisymmetric vortex states from Sec. II. The vortex phase windings are initialized by setting the phases $\phi_j$ as  described in Sec. III. In the following, we present simulation results for three representative initial states $\left(\frac{1}{2},\,\frac{1}{2}\right)$, $\left(0,1 \right)$ and $\left(\frac{1}{
3},\,\frac{1}{3}\right)$, respectively referred to as the `half-half vortex', `zero-one vortex' and the `third-third vortex'.

\subsection{Half--half vortex initial state}

\begin{figure}
 \centering
 \includegraphics[width=0.65\columnwidth]{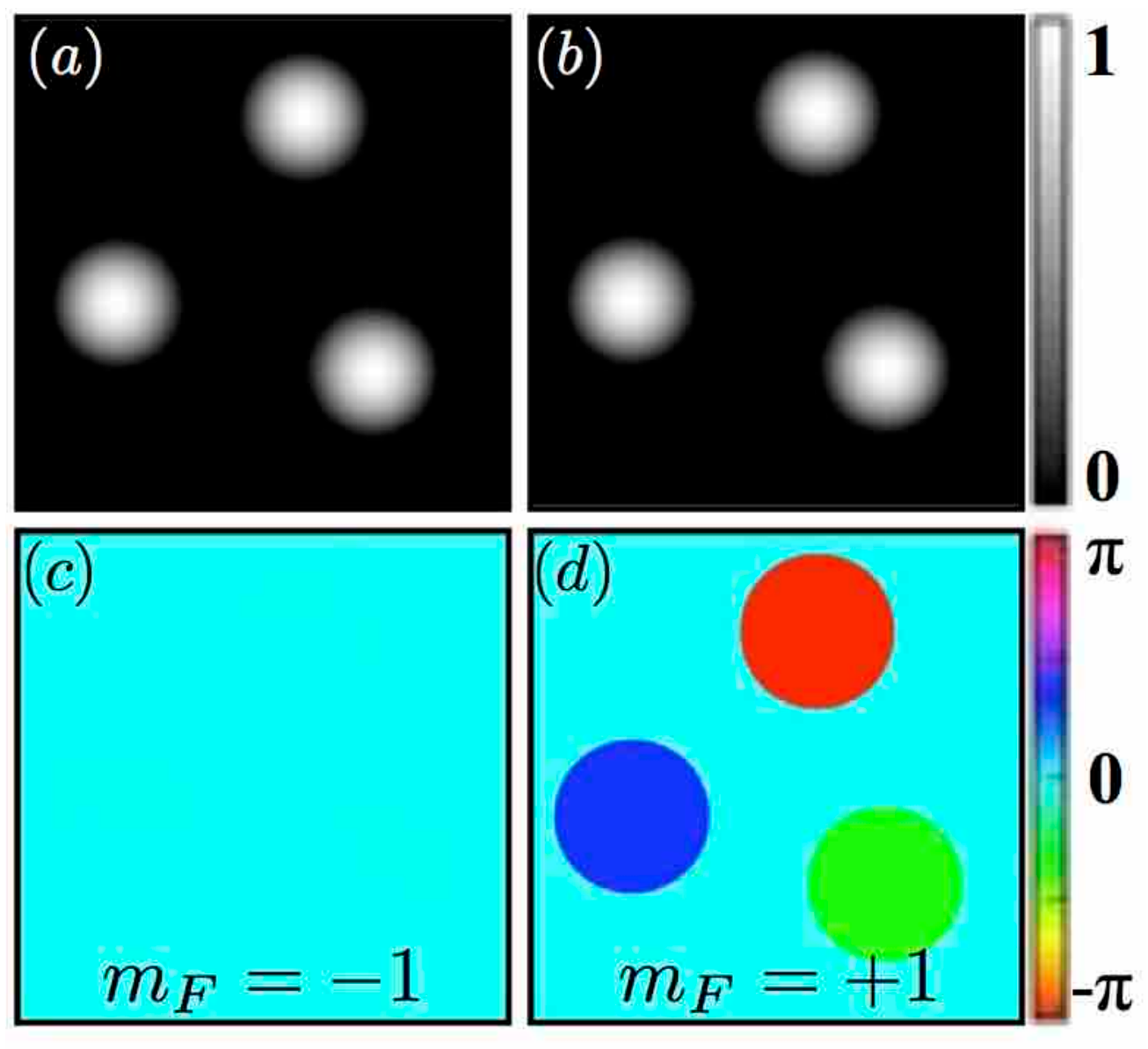}
 \caption{(Color online) Initial state of the $\left(\frac{1}{2},\,\frac{1}{2}\right)$ vortex. (a)-(b) show the probability densities in the $m_F = 1$ and $m_F=-1$ spin states normalized to the maximum density. (c)-(d) show the phases of the two spin states. The field of view of each frame is $(23.4\times23.4)$ $a_\textrm{osc}$.}
 \label{K4}
\end{figure}

The half-half vortex is initialized as the $\left(\frac{1}{2},\,\frac{1}{2}\right)$ state with the spinor phase structure $\left(0,\,1,\,0,\,\exp\left(i\phi\right),\,0\right)$ which in the semi-analytical model of Eq.~\ref{analyticalspinor} corresponds to $\phi_{-1j}={\rm 0}$ and $\phi_{1j}=0,\,\frac{2\pi}{3},\,\frac{4\pi}{3}$ for $j=1,\,2,\,3$, respectively, with the other spin components empty. 
The initial probability density and phase map for the two non-zero population hyperfine spin states is presented in Fig.~\ref{K4} showing the $2\pi$ phase winding across the three condensate fragments in the $m_F=+1$ spin component. Both the magnetization and spin singlet pair amplitude density are initially zero.

\begin{figure*}[htb]
 \centering
 \includegraphics[width=2\columnwidth]{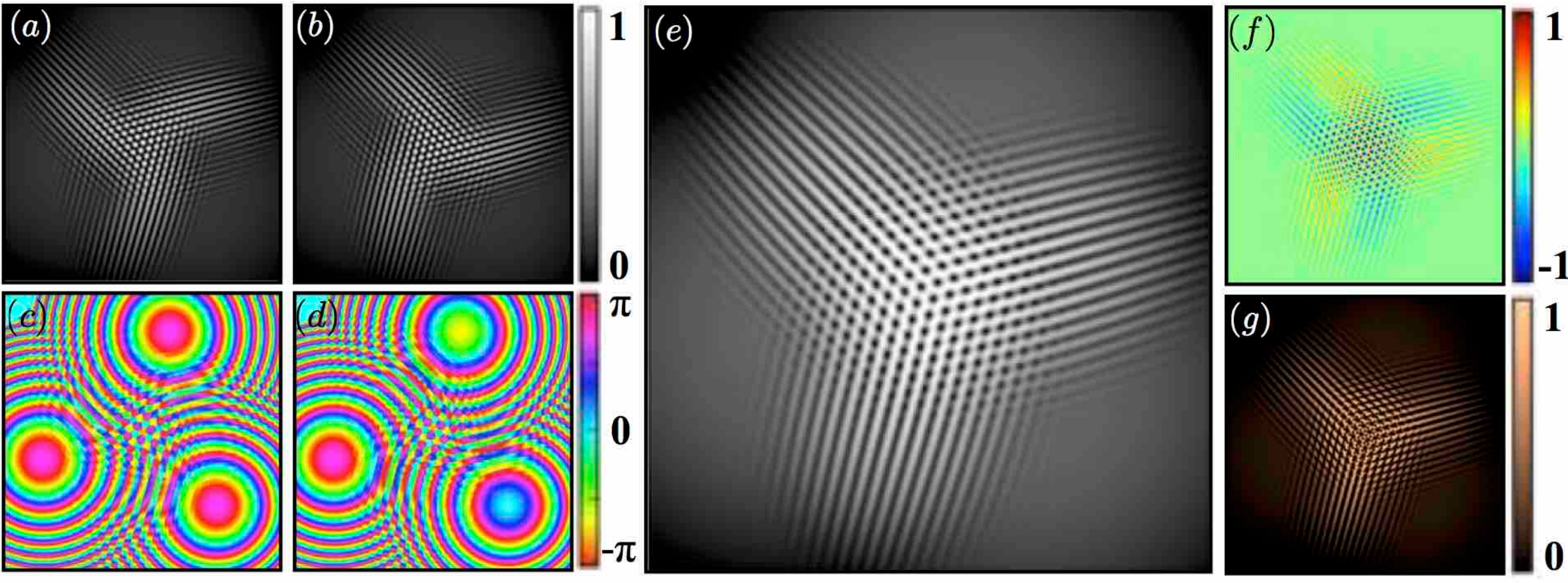}
 \caption{(Color online) The lattice for the $\left(\frac{1}{2},\,\frac{1}{2}\right)$ vortex initial state from a Gross--Pitaevskii simulation. (a)-(d) show the probability density of the $m_F = -1$ and $m_F=1$ spin states normalized to the maximum density and the corresponding phases. (e)-(g) show the total particle density, magnetization density and the spin singlet pair amplitude density. The field of view of each frame is $(23.4\times23.4)$ $a_\textrm{osc}$ and the images are for a time $\tau = 0.55$ $1/\omega$ after the trapping potentials are switched off. See also supplemental Movie S1-S14 \cite{supplemental}.}
 \label{K5}
\end{figure*}

After all external potentials are removed, by setting $V_\textrm{ext}=0$, the three condensate fragments collide and a honeycomb lattice is formed in the condensate interior of each spin state, Fig.~\ref{K5}(a-d), while the exterior regions, where initially only two of the three condensate fragments have collided, are dominated by interference fringes. The vortex lattice consists of the three fractional vortex types $\left(-\frac{1}{2},\,\frac{1}{2}\right)$, $\left(\frac{1}{2},\,\frac{1}{2}\right)$ and $\left(0,\,-1\right)$ as referenced in Table ~\ref{table_initial_states} and shown in Fig.~\ref{K9}(a). The total particle density $n(\gvec{r})$, magnetization density $M(\gvec{r})$ and the spin singlet pair amplitude density $|A(\gvec{r})|$ are shown in Fig.~\ref{K5}(e-g), respectively. The densities in the two populated spin components develop prominent spiral arms with opposite chirality, preserving the three-fold symmetry in the total density.

Figure \ref{K6}(a-c) shows an expanded view of Fig.~\ref{K5}(e-g) local to the trap center. These frames should be compared with the respective frames Fig.~\ref{K6}(d), (e) and (f) showing the corresponding densities calculated using the semi-analytical model. From such comparison it is evident that the semi-analytical non-interacting model is in good agreement with the full Gross--Pitaevskii simulation, as far as the predicted lattice structure is concerned, with the produced vortex types being identical. 

The $\left(\frac{1}{2},\,\frac{1}{2}\right)$ initial state contains equal population of atoms in both spin states $m_F=\pm1$. Therefore the magnetic cores of the half-half type vortices have a magnetization of equal magnitude but opposite sign as shown in Fig. \ref{K6}(b). The $\left(0,\,-1\right)$ vortex has a zero particle density core structure and consequently zero magnetization and spin singlet amplitude. In contrast, the particle density at the cores of half-half type vortices is non-zero. The magnetization density in Fig. \ref{K6}(b) shows a honeycomb pattern of maxima and minima, coinciding with the locations of the $\left(\frac{1}{2},\,\frac{1}{2}\right)$ and $\left(-\frac{1}{2},\,\frac{1}{2}\right)$ vortex cores, respectively. From both the magnetization and the spin singlet pair amplitude we note that the cores of the half-half type vortices have a triangular core structure in comparison to the circular structure of the $\left(0,\,-1\right)$ vortex. The spin singlet pair amplitude density 
displays distinctly different lattice structure when contrasted with the particle and magnetization densities.
For the phase and particle density information for each occupied spin state see the movies S1-S14 in the Supplemental Material \cite{supplemental}.

\subsection{Zero--one vortex initial state}

Consider next a condensate with three spin components populated with condensate particles. For this example we choose the $\left(0,\,1\right)$ vortex initial state, which leads to the ABC stacking of vortex lattices. The initial state is initialized with the spinor phase structure $\left(\frac{i}{2}\exp\left(-2i\phi\right),\,0,\,\frac{\sqrt{2}}{2},\,0,\,\frac{i}{2}\exp\left(2i\phi\right)\right)$ which, in the semi-analytical model of Eq.~\ref{analyticalspinor} corresponds to $\phi_{-2j}= 0,\,\frac{8\pi}{3},\,\frac{4\pi}{3}$, $\phi_{0j}=0$ and $\phi_{2j}=0,\,\frac{4\pi}{3},\,\frac{8\pi}{3}$ for $j = 1,\,2,\,3$, with the other spin components empty. The initial particle density and phase map for the three non-zero population spin states is presented in Supplemental Material, showing the $\pm 2\pi$ phase winding across the condensate fragments in the $m_F=+2$ and $m_F=-2$ spin components.

The condensate collision from the $\left(0,\,1\right)$ vortex initial state, see supplemental movies S15-S32 \cite{supplemental}, produces the lattice vortices $\left(0,\,\frac{1}{2}\right)$, $\left(\frac{1}{2},\,-\frac{1}{4}\right) + \left(-1,\,0\right)$ and $\left(-\frac{1}{2},\,-\frac{1}{4}\right)+\left(1,\,0\right)$ as presented in Table ~\ref{table_initial_states} and the vortex diagram Fig.~\ref{K9}(b). As is evident from Fig.~\ref{K9}(b) the last two vortices cannot be described as single axisymmetric vortices, rather we interpret them to be two-vortex superposition states of the $\left(\frac{1}{2},\,-\frac{1}{4}\right)$ , $\left(-\frac{1}{2},\,-\frac{1}{4}\right)$, $\left(1,\,0\right)$ and $\left(-1,\,0\right)$  vortices defined in Table~ \ref{table_vortex_selection}.
The core of the $\left(0,\,\frac{1}{2}\right)$ vortex has $M=0$ and non-zero particle density. Both $\left(\frac{1}{2},\,-\frac{1}{4}\right) + \left(-1,\,0\right)$ and $\left(-\frac{1}{2},\,-\frac{1}{4}\right)+\left(1,\,0\right)$ have magnetized cores with $M=1$ and $M=-1$ respectively, and both have a non-zero particle density at the vortex core.

\begin{figure}[ht]
 \centering
 \includegraphics[width=\columnwidth]{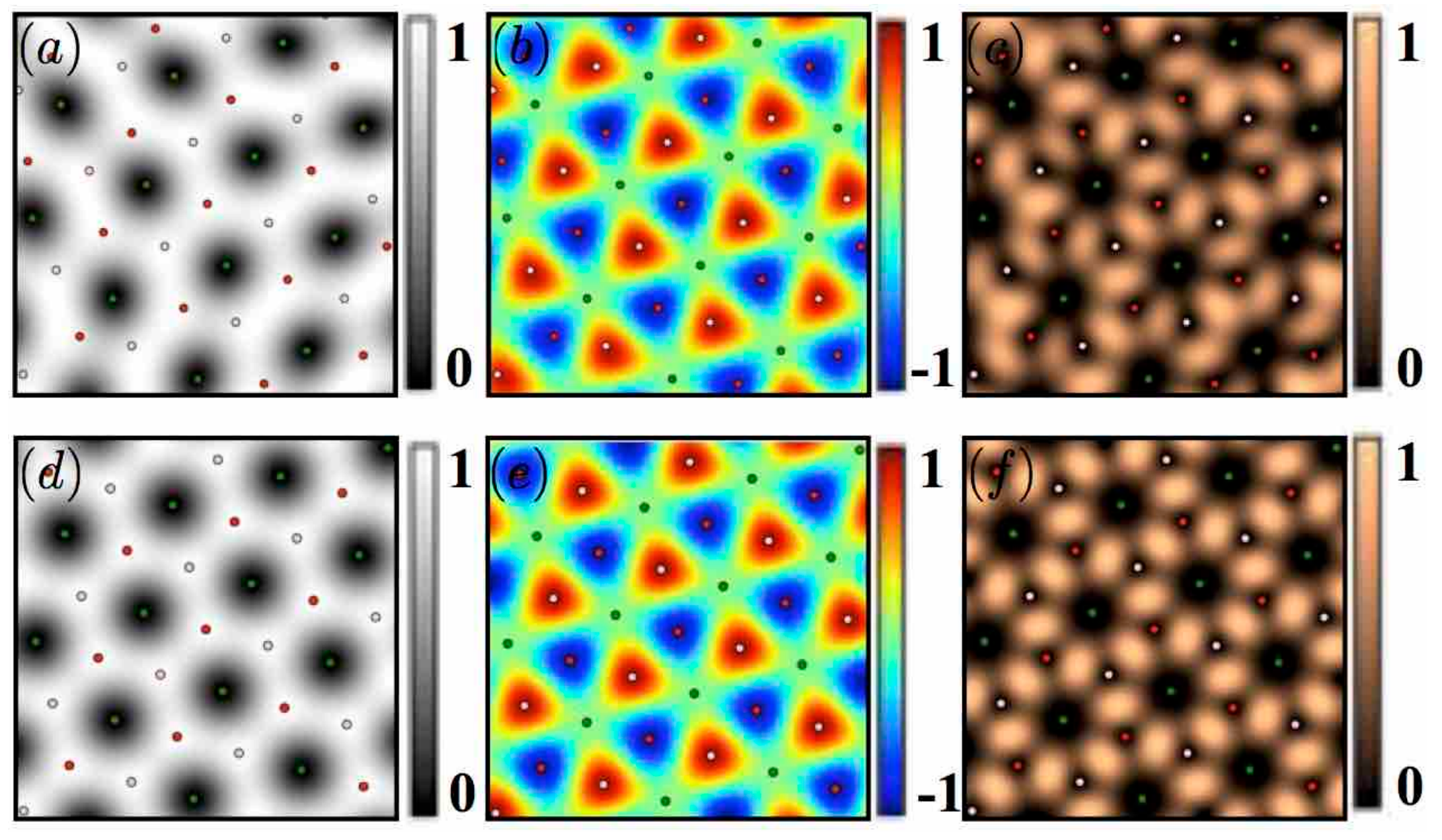}
 \caption{(Color online) An expanded view, local to the trap center, of the lattice for the $\left(\frac{1}{2},\,\frac{1}{2}\right)$ half-half vortex initial state. (a)-(c) The Gross--Pitaevskii simulation for the total particle density, magnetization density and the spin singlet pair amplitude density for a time $\tau = 0.55$ $1/\omega$ after the trapping potentials are switched off. (d)-(f) The corresponding observables calculated from the semi-analytical spinor in Eq.~\ref{analyticalspinor} where $\phi_{-1j}=0$ and $\phi_{1j}=0,\,\frac{2\pi}{3},\,\frac{4\pi}{3}$ for $j=1,\,2,\,3$, with the other spin components empty. The locations of the $\left(\frac{1}{2},\,\frac{1}{2}\right)$, $\left(-\frac{1}{2},\,\frac{1}{2}\right)$ and $\left(0,\,-1\right)$ lattice vortices are denoted by coloured dots of pink, red, and green, respectively. The field of view of each frame is $(2.3\times2.3)$ $a_\textrm{osc}$. For the corresponding movies S12-14, see Supplemental Material \cite{supplemental}.}
 \label{K6}
\end{figure}

An expanded view, local to the trap center, to the vortex lattice structures in each of the three observables is shown in Fig.~\ref{K7} while full lattices for each spin state are presented in the Supplemental Material \cite{supplemental}. As shown in Fig.~\ref{K7}(b) magnetization density emerges due to the nucleation of vortices with magnetic core structures which changes the topology of the condensate. The total particle density, see Fig.~\ref{K7}(a), does not vanish anywhere although the magnetization density remains nearly identical to that in Fig.~\ref{K6}(e). The spin singlet pair amplitude density in Fig.~\ref{K7}(c) reveals the asymmetry between the initial state vortices and those spawned by the condensate collision.

Comparing Fig.~\ref{K7}(a-c) with Fig.~\ref{K7}(d-f), the $\left(0,\,\frac{1}{2}\right)$, $\left(\frac{1}{2},\,-\frac{1}{4}\right)$ , $\left(-\frac{1}{2},\,-\frac{1}{4}\right)$, $\left(1,\,0\right)$ and $\left(-1,\,0\right)$  vortices have nucleated in the spinor wavefunction in complete agreement with our semi-analytical model. The triangular structure of the spin singlet pair amplitude density differs from the semi-analytical prediction of a snow-flake pattern showing it to be more sensitive to the local detail of the spinor order parameter.
 
\begin{figure}[ht]
 \centering
 \includegraphics[width=\columnwidth]{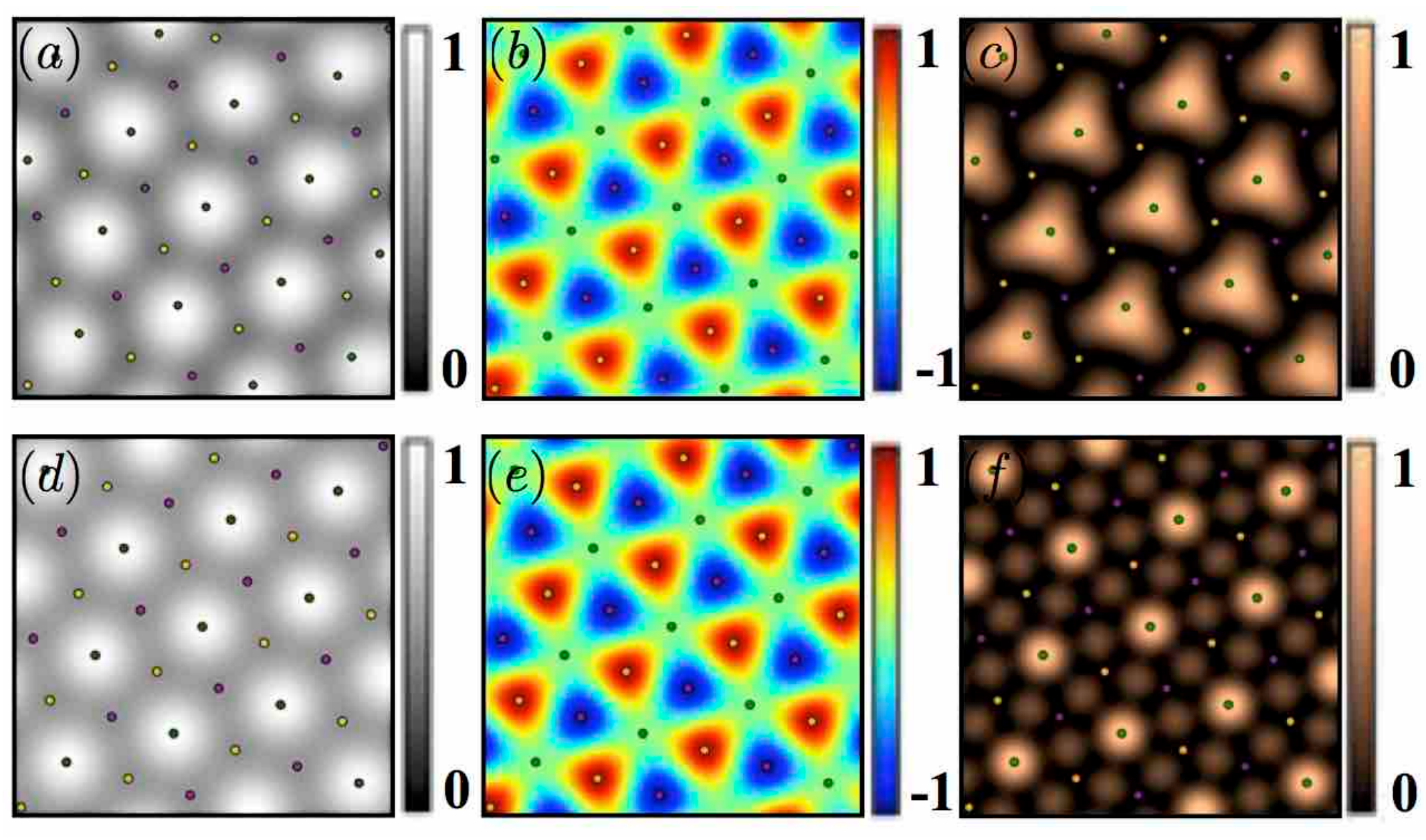}
 \caption{(Color online) An expanded view, local to the trap center, of the lattice for the $\left(0,\,1\right)$ zero-one vortex initial state. (a)-(c) The numerically calculated total particle density, magnetization density and the spin singlet pair amplitude density for a time $\tau = 0.55$ $1/\omega$ after the trapping potentials are switched off. (d)-(f) The corresponding observables calculated with the semi-analytical spinor in Eq.~\ref{analyticalspinor} where $\phi_{-2j}=0,\,\frac{8\pi}{3},\,\frac{4\pi}{3}$, $\phi_{0j}=0$ and $\phi_{2j}=0,\,\frac{4\pi}{3},\,\frac{8\pi}{3}$ for $j=1,\,2,\,3$, respectively, with the other spin components empty. The locations of the $\left(0,\,\frac{1}{2}\right)$, $\left(\frac{1}{2},\,-\frac{1}{4}\right) + \left(-1,\,0\right)$ and $\left(-\frac{1}{2},\,-\frac{1}{4}\right)+\left(1,\,0\right)$ lattice vortices are denoted by coloured dots of green, yellow and purple respectively. The field of view of each frame is $(2.3\times2.3)$ $a_\textrm{osc}$. See also movies 
S30-32 in Supplemental Material \cite{supplemental}.}
  \label{K7}
\end{figure}

\subsection{Third--third vortex initial state}

\begin{figure}[!h]
 \centering
 \includegraphics[width=\columnwidth]{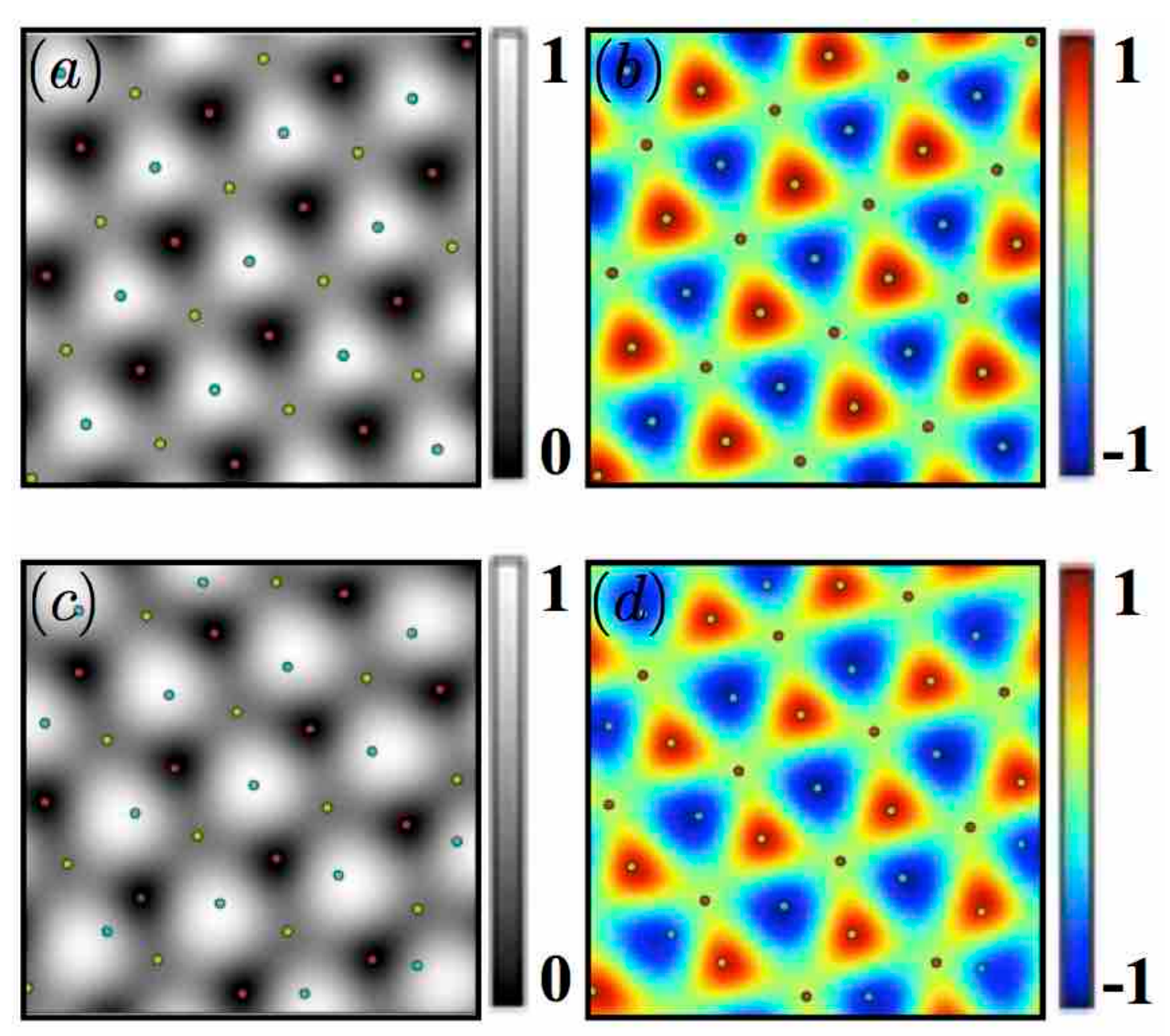}
 \caption{(Color online) An expanded view, local to the trap center, of the lattice for the $\left(\frac{1}{3},\,\frac{1}{3}\right)$ third-third vortex initial state. (a)-(b) The Gross--Pitaevskii calculation for total particle density and  magnetization density for a time $\tau = 0.55$ $1/\omega$ after the trapping potentials are switched off. (c)-(d) The corresponding observables calculated with the semi-analytical spinor in Eq.~\ref{analyticalspinor} where $\phi_{-1j}=0$ and $\phi_{2j}=0,\,\frac{2\pi}{3},\,\frac{4\pi}{3}$ for $j=1,\,2,\,3$, with the other spin components empty. The locations of the $\left(\frac{1}{3},\,\frac{1}{3}\right)$, $\left(-\frac{2}{3},\,\frac{1}{3}\right)$  and $\left(\frac{1}{3},\,-\frac{2}{3}\right)$ lattice vortices are denoted by coloured dots of cyan, yellow and brown respectively. The field of view of each frame is $(2.7\times2.7)$ $a_\textrm{osc}$.}
  \label{K8}
\end{figure}

The third-third vortex initial state consists of the $\left(\frac{1}{3},\,\frac{1}{3}\right)$ vortex with the spinor phase structure $\left(0,\,1,\,0,\,0,\,\exp\left(i\phi\right)\right)$ which in the semi-analytical model of Eq.~\ref{analyticalspinor} corresponds to $\phi_{-1j}={\rm 0}$ and $\phi_{2j}=0,\,\frac{2\pi}{3},\,\frac{4\pi}{3}$ for $j=1,\,2,\,3$, with the other spin components empty. The initial probability densities and phases of the two occupied spin states are shown in Supplemental Material \cite{supplemental}. The AB stacking produces the lattice vortices $\left(\frac{1}{3},\,\frac{1}{3}\right)$, $\left(-\frac{2}{3},\,\frac{1}{3}\right)$  and $\left(\frac{1}{3},\,-\frac{2}{3}\right)$ as presented in Table ~\ref{table_initial_states} and Fig.~\ref{K9}(c). The vortex lattices in each spin state are also presented in Supplemental Material. Importantly, this simulation confirms that, as anticipated, the $\left(\frac{1}{3},\,\frac{1}{3}\right)$ initial state does indeed create a lattice of non-Abelian vortices and antivortices. 
\\
The particle densities at the cores of the $\left(\frac{1}{3},\,\frac{1}{3}\right)$, $\left(-\frac{2}{3},\,\frac{1}{3}\right)$  and $\left(\frac{1}{3},\,-\frac{2}{3}\right)$ vortices are $n=\frac{2}{3} n_\textrm{max}$, $\frac{1}{3} n_\textrm{max}$ and $0$ respectively, where $ n_\textrm{max}$ is the peak total particle density. Thus the $\left(\frac{1}{3},\,-\frac{2}{3}\right)$ vortex has a zero particle density core while the $\left(\frac{1}{3},\,\frac{1}{3}\right)$ and $\left(-\frac{2}{3},\,\frac{1}{3}\right)$  vortex cores have dimensionless magnetizations $-\frac{2}{3}$ and $\frac{2}{3}$, respectively. An expanded view of the lattice structures present in the total particle density and magnetization density is shown in Fig.~\ref{K8}(a-b), while the spin singlet pair amplitude density is zero across all space. Note that the vortex cores in both of the non-zero observables have a prominent triangular structure. The vortex lattices and the total density and 
magnetization density structures are indistinguishable from those predicted by the semi-analytical model.
For the movies corresponding to the evolution of the system in each observable see movies S33-S44 in Supplemental Material \cite{supplemental}.

Based on these three examples, it is clear that the multi-wave interference technique can be used to deterministically produce desired vortex lattice topologies in spinor Bose--Einstein condensates. 

\subsection{Route to non-Abelian quantum turbulence}

\begin{figure*}[htb]
 \centering
 \includegraphics[width=1.8\columnwidth]{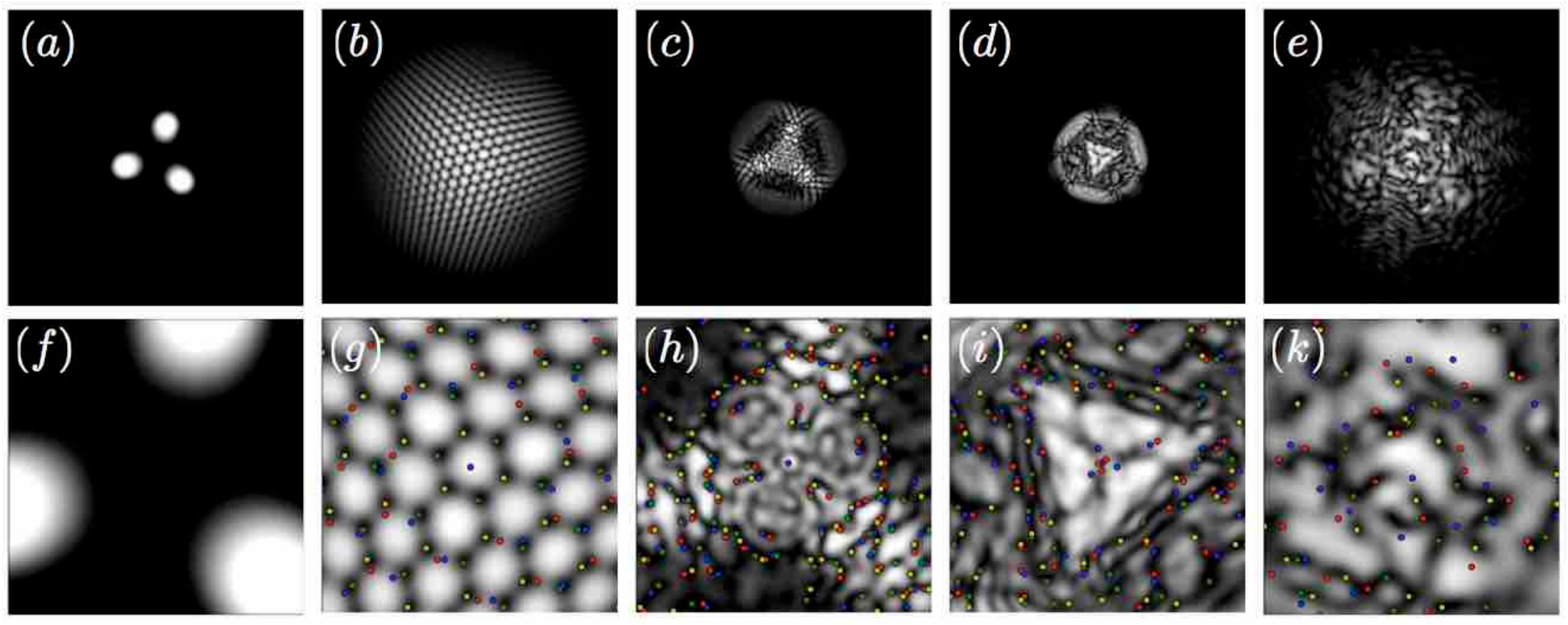}
 \caption{(Color online) Emergence of non-Abelian two-dimensional quantum turbulence showing a selection of snapshots from the $\left(\frac{1}{3},\,\frac{1}{3}\right)$ initial state simulation. (a)-(e): the total probability density of the condensate including the initial state, lattice and turbulent regime. The field of view of each frame is $(29.2\times29.2)$ $a_\textrm{osc}$. (f)-(k): the same frames in (a)-(e) but with a field of view of $(2.92\times2.92)$ $a_\textrm{osc}$. The coloured dots of blue, red, green and yellow denote the vortices (blue/green) and antivortices (red/yellow) in the $m_F=2$ and $m_F=-1$ spin states, respectively. See also Movie S33-S44 in Supplemental Material \cite{supplemental}.}
 \label{K10}
\end{figure*}

As shown in the previous subsection, the $\left(\frac{1}{3},\,\frac{1}{3}\right)$ third-third vortex initial states can be used for generating vortex lattices of non-Abelian vortices and antivortices. Keeping the global harmonic trapping potential turned on and only switching off the triple-well potential will cause the condensate to undergo breathing mode oscillations in the harmonic trap. Figure~\ref{K10}(a)-(e) shows snapshots of such a simulation and the frames (f)-(k) show the enlarged images zoomed to the trap centre. Despite the initial three-fold symmetry, the chaotic dynamics of the vortices rapidly leads to the loss of such symmetry and a transition to turbulence as shown in (c)-(d). A movie of the full simulation is included in the Supplemental Material \cite{supplemental}. 

The instability of the $\left(\frac{1}{3},\,-\frac{2}{3}\right)$ vortex becomes apparent in the movie at $\tau = 0.50$ $1/\omega$ when it splits into an antivortex and a vortex in the $m_F = 2$ and $m_F = -1$ hyperfine spin states, respectively. The uncoupled antivortex typically annihilates with the $\left(\frac{1}{3},\,\frac{1}{3}\right)$ vortex, though both vortices later reform to produce scalar vortex bound states with the $\left(-\frac{2}{3},\,\frac{1}{3}\right)$  and the uncoupled vortex, respectively. The $\left(\frac{1}{3},\,\frac{1}{3}\right)$ vortex at the origin is the only exception to this annihilation process. The decoupling of the $\left(\frac{1}{3},\,-\frac{2}{3}\right)$ vortex replaces the interleaved triangular lattice with a hexagonal lattice of scalar vortices.

We have also performed preliminary fully three-dimensional calculations for the third-third vortex initial states and have verified that the non-Abelian vortex lattices are also produced in this case. In three-dimensional systems the Crow instability \cite{Berloff2001a,Simula2011b} of vortices and anti-vortices leads to the generation of Kelvin waves \cite{PhysRevA.69.043617,PhysRevB.77.060502,PhysRevLett.101.020402}. The growth of Kelvin waves may trigger the vortex collisions leading to the formation of rung vortex networks and three-dimensional non-Abelian quantum turbulence.

\section{Discussion}

We have studied computationally the generation of quantized vortex lattices and quantum turbulence in spin-2 vector Bose--Einstein condensates by simulating collisions of three condensate fragments. We have shown that the structure of the resulting honeycomb vortex lattices can be predicted by modeling each of the spinor wavefunction components independently in terms of linear superposition of three  waves. The lattice states thus produced correctly predict the structure of fractional-vortex lattices observed in full simulations of the spinor Gross--Pitaevskii equation. 

We have shown that using realistic initial state preparation, honeycomb lattices of non-Abelian vortices and antivortices can be produced using three wave interference technique. This is anticipated to open a route to experimental studies of non-Abelian quantum turbulence in vector Bose--Einstein condensates. Despite the relatively short life-times of the $F=2$ Bose--Einstein condensates, the dynamical method presented for creating the honeycomb vortex lattices and their subsequent decay to turbulence should allow sufficiently long time scales for observations of non-Abelian quantum turbulence to be made. The resulting vortex configurations could potentially be observed using the vortex gyroscope imaging method \cite{Powis2014a} in combination with Stern--Gerlach imaging. 

The main simulations presented here were performed using two-dimensional systems and are ideally suited for further studies of two-dimensional non-Abelian quantum turbulence. Applying the three-wave interference technique for generating quantum turbulence to three-dimensional condensates should allow non-Abelian vortex lines to collide generating rung networks and three-dimensional non-Abelian quantum turbulence. There remain many open questions in this context including: can evaporative heating of fractional vortices lead to the emergence of Onsager vortices of non-Abelian kind \cite{PhysRevLett.112.145301,Simula2014a}; is two-dimensional non-Abelian quantum turbulence characterized by a non-Kolmogorov power law of the incompressible kinetic energy spectrum; and does magnetization cascade emerge in these systems alongside incompressible kinetic energy and enstrophy cascades?  These important questions are left as topics for further studies.

\begin{acknowledgments}
TS acknowledges financial support from the Australian Research Council via Discovery Project DP130102321.
\end{acknowledgments}

\bibliography{master}
\end{document}